\documentclass[10pt]{article}

\usepackage{epsfig}
\usepackage{a4wide}
\usepackage{latexsym}
\usepackage{amssymb}
\usepackage{amsmath}
\usepackage{graphics}

\begin{document}

\def\O{{\cal O}}
\def\N{{\cal N}}
\def\>t{>_{\scriptscriptstyle{\rm T}}}
\def\enu{\epsilon_\nu}
\def\pint{\int{\d^3p\over(2\pi)^3}}
\def\gint{\int[\D g]\P[g]}
\def\hxi{\hat x_i}
\def\hatx{{\bf \hat x}}
\def\d{{\rm d}}
\def\e{{\bf e}}
\def\x{{\bf x}}
\def\X{{\bf X}}
\def\0x{\x^\smalze}
\def\sperpx{{x_\perp}}
\def\sperpk{{k_\perp}}
\def\sbperpk{{{\bf k}_\perp}}
\def\sbperpx{{{\bf x}_\perp}}
\def\perpx{{x_{\rm S}}}
\def\perpk{{k_{\rm S}}}
\def\bperpk{{{\bf k}_{\rm S}}}
\def\bperpx{{{\bf x}_{\rm S}}}
\def\r{{\bf r}}
\def\q{{\bf q}}
\def\zr{{\bf z}}
\def\R{{\bf R}}
\def\A{{\bf A}}
\def\v{{\bf v}}
\def\u{{\bf u}}
\def\w{{\bf w}}
\def\U{{\bf U}}
\def\cm{{\rm cm}}
\def\l{{\bf l}}
\def\sec{{\rm sec}}
\def\Ckol{C_{Kol}}
\def\flux{\bar\epsilon}
\def\zq{{\zeta_q}}
\def\b{b_{kpq}}
\def\bun{b^{\scriptscriptstyle (1)}_{kpq}}
\def\bdu{b^{\scriptscriptstyle (2)}_{kpq}}
\def\z0q{{\zeta^{\scriptscriptstyle{0}}_q}}
\def\smalF{{\scriptscriptstyle {\rm F}}}
\def\smalze{{\scriptscriptstyle (0)}}
\def\smalel{{\scriptscriptstyle (l)}}
\def\smalun{{\scriptscriptstyle (1)}}
\def\smaldu{{\scriptscriptstyle (2)}}
\def\smaltr{{\scriptscriptstyle (3)}}
\def\smalqu{{\scriptscriptstyle (4)}}
\def\smalci{{\scriptscriptstyle (5)}}
\def\smalL{{\scriptscriptstyle{\rm L}}}
\def\smalT{{\scriptscriptstyle {\rm T}}}
\def\smalE{{\scriptscriptstyle{\rm E}}}
\def\smal1n{{\scriptscriptstyle (1,n)}}
\def\smaln{{\scriptscriptstyle (n)}}
\def\smalA{{\scriptscriptstyle {\rm A}}}
\def\shell{{\tt S}}
\def\ball{{\tt B}}
\def\nav{\bar N}
\def\micron{\mu{\rm m}}
\font\brm=cmr10 at 24truept
\font\bfm=cmbx10 at 15truept

\baselineskip 0.7cm

\centerline{\brm Transport properties of heavy particles}
\centerline{\brm in high Reynolds number turbulence} 
\vskip 20pt
\centerline{Piero Olla}
\vskip 5pt
\centerline{ISAC-CNR}
\centerline{Sezione di Lecce}
\centerline{73100 Lecce Italy}
\vskip 20pt

\centerline{\bf Abstract}
\vskip 5pt
The statistical properties of heavy particle trajectories in high Reynolds numbers turbulent flows
are analyzed. Dimensional analysis assuming Kolmogorov scaling is compared with the result of 
numerical simulation using a synthetic turbulence advecting field. The
non-Markovian nature of the fluid velocity statistics along the solid particle
trajectories is put into evidence,
and its relevance in the derivation of Lagrangian transport models is 
discussed.

\vskip 2cm
\vfill\eject

\centerline{\bf I. Introduction}
\vskip 5pt
A heavy particle in a turbulent flow moves along a trajectory, which does not coincide with that
of the fluid parcel where it lied initially \cite{yudine59}. This phenomenon is controlled by 
the particle trajectory relaxation time $\tau_S$, which, in general,
is a function of the difference between the velocities $\v$ and $\u$ respectively of the 
particle and of the fluid at the particle position. To lowest order in the particle Reynolds
number $Re_p$, $\tau_S$ is given by the Stokes time. (For a spherical particle, the particle 
Reynolds number and the Stokes time are defined respectively as $Re_p=|\u-\v|a/\nu$ and 
$\tau_S=\frac{2a^2}{9\nu}|1-\rho/\rho_0|$ with $\rho$ and $\rho_0$ the particle and fluid 
density, $\nu$ the fluid kinematic viscosity and $a$ the particle diameter).
Divergence of fluid and particle trajectories receive contribution both from external forces
(e.g. gravity) acting on the particle and acceleration along a fluid trajectory; $\v$ obeys 
therefore, to lowest order in $Re_p$ and $\rho_0/\rho$, the linear equation:
$$
\dot\v=\tau_S^{-1}(\u-\v)+{\bf g}
\eqno(1)
$$
with $(\rho-\rho_0){\bf g}$ the force acting on the particle.
Accelerations produced by non-uniformity or non-stationarity of the mean flow may be 
reabsorbed into the term ${\bf g}$, after redefinition of $\u$ and $\v$, through 
a Galilean shift to the reference frame of the mean flow.

Starting from the work of Csanady \cite{csanady63}, much of the interest has been 
in the role of gravity, with the conclusion that its main effect on turbulent 
diffusion is a renormalization of the correlation time. A particle, in 
the presence of an external force, will drift with respect to the fluid with an 
average velocity $\tau_S{\bf g}$, a phenomenon known under the name of trajectory crossing, 
and this will cause a shorter time of permanence of a particle in a correlated region and a 
Lagrangian correlation time shorter than that of a fluid parcel.

This renormalization has been used for the derivation of Lagrangian
models for heavy particles dispersion \cite{sawford91}, but it is not clear whether a Markovian 
hypothesis for the velocity is indeed applicable to the case of a heavy particle.
This problem, associated with the small turbulent scales being seen as frozen by the drifting
particle, was not present in \cite{pismen78,nir79}, who dealt essentially with a turbulent flow
with only large scales present. The same turbulent regime was considered in \cite{reeks77},
based on an expansion around the case of an infinite inertia particle.
In the Eulerian approach of \cite{shih86}, the same problem was 
avoided working in the limit of Stokes time shorter than the Kolmogorov time, which is the 
same assumption on which the classical theory by Tchen is based \cite{hinze}.

A different approach is to model fluid and solid particle trajectories on the same ground. 
The result of this line of reasoning is a
class of Lagrangian models, of which, one of the prototypes is described in \cite{desjonqueres88,
berlemont90}. The basic idea is to couple Eqn. (1) (or its more complex counterpart valid for 
arbitrary $Re_p$) with the equations for the solid particle trajectory and with the model
equation for the trajectory of the fluid parcel where the particle lied initially. In high 
Reynolds numbers, one resorts typically to models which are Markovian in the Lagrangian
velocity of the fluid parcel $\u^\smalF(t)$ (see e.g. \cite{thomson87}). In this regime, the 
model equations would read, in the absence of a mean flow:
$$
\begin{cases}
\dot\x=\v;
\qquad
\dot\x^\smalF=\u^\smalF
\\
\dot\u^\smalF
={\bf a}(\u^\smalF,\x^\smalF)+(C_0\bar\epsilon(\x^\smalF))^\frac{1}{2}{\boldsymbol{\xi}};
\qquad
\langle\xi_i(t)\xi_j(0)\rangle=\delta_{ij}\delta(t)
\end{cases}
\eqno(2)
$$
where $a(\u^\smalF,\x^\smalF)$ is of the order of $u^\smalF/\tau_L(\x^\smalF)$, with 
$\tau_L(\x^\smalF)$ the turbulent integral time, $\bar\epsilon(\x^\smalF)$ is the mean turbulent 
dissipation and $C_0$ is an empirical constant, with values lying between 2 and 7 depending on 
the kind of flow \cite{sawford91a}; $\x^\smalF$ is the coordinate of the fluid parcel
for which initially: $\x^\smalF=\x$.
After a time $\Delta$, whose determination is part of the modelling effort, a new fluid parcel
is considered, thus setting again $\x^\smalF=\x$ and the process is repeated; in the meanwhile, 
when $\x^\smalF\ne\x$, the problem is to calculate from $\u^\smalF$ the fluid velocity $\u$ at 
the solid particle position $\x$. 

This approach is able in principle to deal with both regimes of strong external forcing and
dominant trajectory crossing, and of random trajectory separation produced solely by 
finite inertia and particles being unable to follow the rapid turbulent fluctuations.
Obtaining the fluid velocity at the particle position turns out to be
the difficult part of the problem. In \cite{desjonqueres88,berlemont90}, the difference
$\u-\u^\smalF$ was calculated only at the discrete times $t_n=n\Delta$, with the values of 
$\u-\u^\smalF$ at different $t_n$ assumed uncorrelated.  
However, although there have been other attempts in the direction of Markov modelling
\cite{lu92}, the statistics for $\u$ appears to 
be non-Markovian, in general, both in time and in the separation $\x-\x^\smalF$. 
Only recently \cite{shao95,reynolds00}, has the importance of anomalous scaling for the velocity
along solid particle trajectories been recognized. However, the mechanism for its production was 
only partially discussed and no prediction for the dependence 
of the scaling exponent on the flow parameters was provided.

The problems associated with the non-Markovian nature of the fluid velocity $\u$ are further 
complicated in the case of weak external forces, with turbulent fluctuations and inertia 
dominating the process of trajectory 
separation. Analysis of this problem and of its consequence in terms of constraints on the possible 
form of turbulent transport models will be the subject of this paper.
\vskip 10pt

\centerline{\bf II. Scaling analysis}
\vskip 5pt
One can see a turbulent flow as a superposition of vortices of different size $r$, which, at
high Reynolds numbers, obey Kolmogorov scaling. In terms of the space (Eulerian) longitudinal 
structure function, this means, for $r\ll L$, with $L$ of the order of
the size of the largest eddies:
$$
u_r^2=\langle [u_r^\smalE(\x+\r,t)-u_r^\smalE(\x,t)]^2\rangle
=C_K(\bar\epsilon r)^\frac{2}{3}
=u_L^2\Big(\frac{r}{L}\Big)^\frac{2}{3}
\eqno(3)
$$
with $\u^\smalE(\x,t)$ the Eulerian velocity field, $C_K\simeq 2$ the Kolmogorov 
constant and $u_L^2=\langle [u_i^\smalE]^2\rangle$. In terms of the time (Lagrangian) structure 
function,
for $t\ll\tau_L$:
$$
\langle [u^\smalF_i(t)-u^\smalF_i(0)]^2\rangle =C_0\bar\epsilon t=u_L^2\frac{t}{\tau_L}
\eqno(4)
$$
with $\tau_L$ of the order of the Lagrangian correlation time $T_L$:
$$
T_L=u_L^{-2}\int_0^\infty\d t\langle u^\smalF_i(t)u^\smalF_i(0)\rangle.
\eqno(5)
$$
The quantities $L$, $\tau_L$ and $u_L$ identify integral length, time and velocity scales 
of the flow, and, from Eqns. (3-4), an eddy turnover time can be defined in the standard way:
$$
\tau_r=\tau_L\Big(\frac{r}{L}\Big)^\frac{2}{3}
\eqno(6)
$$
Consider a solid particle, which at time 
$t=0$ is in $\x_0$ moving with speed $\v_0$ and to fix the ideas imagine first that no external 
forces are present. The trajectory separation will be given by:
$$
\Delta\x(t)=\x(t)-\x^\smalF(t)=\int_0^t\d t'[\v(t')-\u^\smalF(t'|\x_0,0)]
\eqno(7)
$$
where $\u^\smalF(t|\x_0,0)$ is the fluid velocity of the fluid parcel, which at time $t=0$
was in $\x_0$. 

Asymptotically, for $\tau_S/\tau_L,\tau_\eta/\tau_S\to 0$, where 
$\eta=\nu^\frac{3}{4}\bar\epsilon^\frac{1}{4}$ is the Kolmogorov length and $\tau_\eta$ is the
associated Kolmogorov time, we can distinguish three phases in the separation process.
The central role is played by the velocity and length scales of the vortices with life-time 
$\tau_S$, namely, for $\tau_L>\tau_S$:  
$$
u_S=u_L(\tau_S/\tau_L)^\frac{1}{2},\qquad
S=L(\tau_S/\tau_L)^\frac{3}{2}= u_S\tau_S;
\eqno(8)
$$ 
$\tau_S$ and $u_S$ will appear to be the time and velocity scale of the velocity 
difference $\u-\v$.

If $\u_0\equiv\u(0)=\u^\smalF(0|\x_0,0)$ is the initial fluid velocity and $|\v_0-\u_0|\ll u_S$,
there will be an 
initial phase, in which the particle sees the turbulent field as frozen and the separation
process is ballistic:
$$
\Delta\x(t)\simeq (\v_0-\u_0)t.
\eqno(9)
$$
In fact, if 
$\Delta x(t)$ is so small that vortices of size $\Delta x(t)$ have characteristic 
velocity $u_{\Delta x(t)}\ll |\v_0-\u_0|$, it will also be $\frac{\Delta x(t)}{|\u^\smalF-\v|}
\ll\tau_{\Delta x(t)}$, i.e. the crossing time of the vortex by the solid particle
will be much shorter than the eddy turnover time of that vortex. The contribution from this vortex 
to $\Delta x$ will thus be $u_{\Delta x(t)}\frac{\Delta x(t)}{|\u^\smalF-\v|}\ll\Delta x(t)$
and then $\Delta x(t)\simeq |\u_0-\v_0|t$. This phase will last until $u_{\Delta x(t)}\sim
|\u_0-\v_0|$, when the contribution from the vortices begins to be important. For the
Kolmogorov spectrum described by Eqn. (2), this leads to a cross-over time:
$$
\tau_B\sim C_K^{-\frac{3}{2}}\bar\epsilon^{-1}|\v_0-\u_0|^2,
\eqno(10)
$$
and for $|\v_0-\u_0|\ll u_S$, it will also be $\tau_B\ll\tau_S$.

In the next phase, for $\tau_B\ll t\ll\tau_S$, the separation process will receive
two contributions: one from the velocity difference across a vortex of size $\Delta x(t)$,
the other from the fact that the solid particle is unable to follow the fluid on time
scales much shorter that $\tau_S$. 
From the relaxation equation nature of Eqn. (1), these contributions 
are both of the same order, and the end result is that the particles 
will separate with a relative velocity of the order of $u_{\Delta x(t)}$. This leads to
a process of Richardson diffusion $\langle\Delta x^2(t)\rangle\propto t^3$, although with a 
coefficient different from that occurring 
in the case of a pair of fluid particles. Now, for most choices of initial conditions:
$|\v_0-\u_0|\sim u_S$, the ballistic phase will end at $t\sim\tau_B\sim\tau_S$ 
and the Richardson diffusion phase will be absent. The same will occur
when $|\v_0-\u_0|>u_S$; the ballistic phase will end when the solid particle trajectory 
relaxes, which happens, again, when $\tau_B\sim\tau_S$.

In the end phase, for $t\gg\tau_S$, the solid particle will deviate from the
fluid parcel it crosses at the time $t$, with a rate much smaller than the rate
of separation from the fluid parcel it crossed at time zero. In fact, from Eqn. (1), 
the difference $\v-\u$ will saturate at $t\sim\tau_S$, when its magnitude is $O(u_S)$; 
for $t>\tau_S$, $\v-\u$ will continue to fluctuate with amplitude 
$u_S$ and correlation time $\tau_S$. 
In this end phase, the equation for $\Delta\x$ 
will have the form:
$$
\Delta\dot\x(t)\simeq
\u^\smalF(t|\x(\tau_S),\tau_S)-\u^\smalF(t|\x^\smalF(\tau_S),\tau_S)
\eqno(11)
$$
i.e. Richardson law for the separation of two fluid particles, which, at time $\tau_S$,
were distant $\Delta x(\tau_S)=|\x(\tau_S)-\x^\smalF(\tau_S)|\sim S$. Of course, when 
$\Delta x(t)>L$ (and $t>\tau_L$), the fluctuations at $\x$ and $\x_F$ will decorrelate
and Richardson diffusion will turn into normal diffusion. In the extreme case of very fast
turbulent fluctuations, i.e. $\tau_L<\tau_S$, the Richardson diffusion phase may be absent 
and separation
will cross over directly to a regime of normal diffusion, without passing through the regime 
described by Eqn. (11), with $|\u-\v|$ saturating to $u_L$ instead of to $u_S$.

\vskip 5pt
The picture is complicated a bit by the presence of external forcing on the
particle; this leads to new characteristic scales for velocity, length and time:
$$
u_G=g\tau_S,
\qquad
G= L(g\tau_S/u_L)^3
\quad{\rm and}\quad
\tau_G=\tau_L(g\tau_S/u_L)^\frac{3}{2},
\eqno(12)
$$ 
associated respectively with the drift of the particle in the fluid, and with
the size and lifetime of the eddies having this characteristic velocity. The relative 
importance of external forces and of random acceleration by turbulence is fixed by the 
ratio $u_G/u_S$. Both limit regimes $\tau_S=0,u_G\ne 0$ and $\tau_S>0,g=0$ are physically
realizable, an example of the first one being that of charged buoyant particles 
($\rho=\rho_0$) in a strong electric field.
External forcing will have a negligible effect if $u_G<\min(u_S,u_L)$, i.e., from Eqn. (12), if:
$$
u_L>\frac{g\tau_S}{\min(1,\tau_S/\tau_L)^\frac{1}{2}}
\eqno(13)
$$
Thus, inertia may be dominant, when turbulent fluctuations are so strong and fast that
$u_L>g\tau_S$ and $\tau_L<\tau_S$ (for $\rho/\rho_0\gg 1$, in which case external forces are 
an important factor, this happens typically in the vicinity of obstacles) or otherwise, when
the particles are small and
the Stokes time is short enough: $u_L>g(\tau_S\tau_L)^\frac{1}{2}$.

In the presence of external forces, the difference $\u-\v$ will fluctuate with amplitude $u_S$ and
time scale $\tau_S$, around $\u_G$ instead of around zero. For $u_G\ll u_S$, trajectory
separation develops in the same sequence of phases as in the case without forces; the time
$\tau_G<\tau_S$ identifies only the value of the crossover time $\tau_B$ corresponding to 
$|\u_0-\v_0|\sim u_G$.

When external forcing dominates over the effect of inertia, i.e. when $u_G\gg u_S$, the sequence 
of phases is altered.
The initial phase $t<\tau_B$ remains ballistic in nature. Also the next one, for
$\tau_B\ll t\ll\tau_S$, is not modified; since $gt\ll u_{\Delta x(t)}$, the external force 
contribution is small and the particle separation follows the same pattern {\it a la} 
Richardson as in the $g=0$ case. For $\tau_S\ll t\ll\tau_G$,
however, a new phase will occur: the particle is drifting with velocity $u_G$ and the contribution 
to separation from vortices of size $\Delta x(t)$ is negligible: $\Delta x(t)\ll G$ and 
$u_{\Delta x(t)}\ll u_G$. The end phase of pure Richardson diffusion described by Eqn. (11)
starts only when $t>\tau_G$, so that $u_{\Delta x(t)}>u_G$, but may be absent in the case of  
external forces dominating over turbulence: $u_G>u_L$. 
In this case, also the end phase will be dominated by the
drift, with superimposed a normal diffusion correction produced by eddies at scales $L$. 
While in the absence of gravity, the cross-over to fluid-like behaviors occurred at the 
space scale $S$, in the present case the transition occurs at the scale $G$, which, since
$u_G\gg u_S$, is larger than $S$.

In the three cases of dominant inertia, intermediate (in which
external forces dominate inertia, but are dominated by turbulence), and external forces
dominating over both inertia and turbulence, the separation 
process will develop in the following possible sequences of phases:

\noindent {\bf Dominant inertia} $\tau_G<\tau_S$: Ballistic $(t<\tau_B)\ \rightarrow$ 
Modified Richardson 
$(\tau_B<t<\tau_S)\ \rightarrow$ Richardson $(\tau_S<t<\tau_L)\ \rightarrow$ Normal $(t>\tau_L)$;

\noindent {\bf Intermediate} $\tau_S<\tau_G<\tau_L$: Ballistic 
$(t<\tau_B)\ \rightarrow$ Modified Richardson 
$(\tau_B<t<\tau_S)\ \rightarrow$ Drift $(\tau_S<t<\tau_G)\ \rightarrow$
Richardson $(\tau_G<t<\tau_L)\ \rightarrow$ Normal $(t>\tau_L)$.

\noindent {\bf Dominant External Forces} $\tau_G>\tau_L$: Ballistic $(t<\tau_B)\ \rightarrow$ 
Modified Richardson $(\tau_B<t<\tau_S)\ \rightarrow$ Drift $(t>\tau_S)$

\vskip 10pt
The considerations that have been carried on so far correspond to the intuitive fact that the 
turbulent small scales, i.e. 
those for which $l<\max(S,G)$, are unable to transport the solid 
particles. For $l<S$ this occurs because of inertia; for $S<l<G$, it is the external forcing
produced trajectory crossing which is responsible for this. In the case of the large scales
$l>\max(S,G)$, the contrary occurs: one has $|\u-\v|\sim\max(u_S,u_G)<u_l$ and the deviation 
of the particle trajectory from that of a fluid parcel on the scale of a vortex of size $l$ 
will be $\sim u_{S,G}(\tau_{S,G}\tau_l)^\frac{1}{2}<l$ and the solid particle will behave like 
a fluid one. 

This scale separation of the turbulent fluctuations and the the subdivision in phases of the 
trajectory separation process, has a counterpart in the self-diffusion 
$\langle |\u(t|\x_0,0)-\u_0|^2\rangle$, which is the relevant quantity to determine
the transport properties of the particles $[\u(t|\x_0,0)$ is the fluid velocity sampled by the 
solid particle, which at time $t=0$ was in $\x_0]$.
Following \cite{berlemont90} we can split the velocity
difference $\u(t|\x_0,0)-\u_0$ into the different contributions along, and transversal to fluid 
trajectories.
The first contribution is just the Lagrangian change $\u^\smalF(t|\x_0,0)-\u_0$; the second one 
$\u(t|\x_0,0)-\u^\smalF(t|\x_0,0)=\u^\smalE(\x(t),t)-\u^\smalE(\x^\smalF(t),t)$ is a complex 
quantity feeling both effects of trajectory separation and of the relative diffusion of fluid 
elements. In the ballistic regime, using Eqn. (3), we have:
$$
\u(t|\x_0,0)-\u^\smalF(t|\x_0,0)
\simeq\u^\smalE(\x_0+\v_0t,t)-\u^\smalE(\x_0+\u_0t,t)
$$
$$
\sim u_{|\u_0-\v_0|t}
\sim u_L(|\u_0-\v_0|t/L)^\frac{1}{3}
$$ 
while, from Eqn. (4):
$$
\u^\smalF(t|\x_0,0)-\u_0
\sim \frac{u_Lt}{\tau_L}\ll u_L\Big(\frac{|\u_0-\v_0|t}{L}\Big)^\frac{1}{3}
$$ 
Therefore:
$$
\langle |\u(t|\x_0,0)-\u_0|^2\rangle\sim C_K(\bar\epsilon|\v_0-\u_0|t)^\frac{2}{3}
\sim  C_K(\bar\epsilon u_At)^\frac{2}{3}
\eqno(14)
$$ 
where $t\ll\tau_B$ and $A=S,G$.

In the opposite regime $t\gg\tau_{S,G}$, from Eqn. (11), the transverse separation will be
dominated by
the relative diffusion of fluid elements: 
$\u(t|\x_0,0)-\u^\smalF(t|\x_0,0)\simeq\u^\smalF(t|\x(\tau_S),\tau_S),
-\u^\smalF(t|\x^\smalF(\tau_S),\tau_S)$
while the ''longitudinal'' component remains 
$\u^\smalF(t|\x_0,0)-\u_0\sim (C_0\flux t)^\frac{1}{2}$. 
Also the first component obeys normal scaling, but its presence comes simply from having taken
a wrong initial condition for $\u(t|\x_0,0)\simeq\u^\smalF(t|\x(\tau_S),\tau_S)$.
Said in a different way, the transverse component is not independent of the longitudinal one,
and they sum together in such a way that the total amplitude is equal to that of the longitudinal
piece.  The total velocity variation is therefore:
$$
\langle |\u(t|\x_0,0)-\u_0|^2\rangle\sim C_0\bar\epsilon t
\qquad
\tau_L\gg t\gg\tau_S,\tau_G
\eqno(15)
$$
i.e., for long times and turbulence strong enough to prevent drift dominance at
large $t$, the self-diffusion properties of fluid and solid particles will be the same.

In the middle range $\tau_B\ll t\ll\tau_S$, all the contributions to $\u(t|\x_0,0)-\u_0$ are of the
same order; these are the Lagrangian change $\u^\smalF(t|\x_0,0)-\u_0$, 
the one from Richardson 
diffusion starting at times $\tau_B\ll t$: $\u^\smalF(t|\x(\tau_B),\tau_B)
-\u^\smalF(t|\x^\smalF(\tau_B),\tau_B)$, and the remnant $\u(t|\x(\tau_B),\tau_B)-
\u^\smalF(t|x(\tau_B),\tau_B)$, which is due to trajectory separation.
In fact, $t$ identifies a scale $l$, through the relation $\tau_l=t$, 
which is the size of the vortices contributing most both to $\u(t|\x_0,0)-\u_0$, to Richardson 
diffusion and to the trajectory separation $|\u(t)-\v(t)|\sim u_l$.
Even after elimination of the Richardson diffusion by choosing a different Lagrangian path, 
the difference $\u(t|\x_0,0)-\u_0$ will receive a contribution proportional to $u_l$ due to 
the $O(l)$ displacement between the solid and fluid particle trajectory. This is of the
same order as the Lagrangian piece $\u^\smalF(t|\x_0,0)-\u_0$, associated with the time 
evolution of a size $l$ vortex, which is $O(u_l)$ in the time $t\sim\tau_l$. The end 
result will be again, the normal behavior for velocity self-diffusion, described in 
Eqn. (15), 
but with a renormalized $C_0$ to account for the trajectory separation $|\u(t)-\v(t)|\sim u_l$.
\vskip 10pt

\centerline{\bf III. Transport by a synthetic turbulent field}
\vskip 5pt
One of the difficulties in studying the divergence of solid and fluid particles is that the 
process is affected simultaneously 
by Eulerian and Lagrangian properties of the turbulent flow. To simulate numerically the
trajectory of a single solid particle, one would expect therefore that a whole advecting
field should be generated. Clearly, this is as difficult as generating a high Reynolds number 
turbulent flow by means of a DNS (direct numerical simulation) of the Navier-Stokes equation.
The strategy to bypass this problem, while working in high Reynolds number conditions, has 
been to generate numerically an advecting flow only in the neighborhood of the
solid particle. This flow satisfies a restricted set of properties of real turbulence,
namely, incompressibility, Kolmogorov scaling for the space structure function and normal 
scaling for the Lagrangian time structure function, as described by Eqn. (4).

The flow was generated by means of a superposition of vortices
with the following properties (more details are given in the Appendix):
\newcounter{rom}
\begin{list}
{\roman{rom}}{\usecounter{rom}\setlength{\rightmargin}{\leftmargin}}
\item The vortices are distributed in a sequence of shells: those in shell $n$ have size 
$l\in [2^{-n-1},2^{-n}]$ and are distributed uniformly in the logarithm of scale \cite{olla98}. 
In this way, if $N+1$ is the total number of shells: $0\le n\le N$, and thus 
a Kolmogorov length $\eta=2^{-N-1}$ and an effective
Reynolds number $Re=L/\eta\sim 2^{N+1}$ are introduced in a natural way.
\item The vortices have a finite lifetime $\bar\tau_l=\beta l^\frac{2}{3}$, with $\beta$ a fixed
constant, and are generated in the neighborhood of the particle in such a way that the particle 
lies on the average in the support of only one vortex per shell. The radius of this neighborhood
scales linearly with the shell scale, so that all shells have, on the average, the same number
of vortices. 
\item A vortex in shell $n$ translates
rigidly with speed equal to the value at its center of the velocity field produced by all 
vortices in shells $n'<n$. 
\end{list}
For the kind of analysis of scaling and crossover behaviors we are interested in, two dimensions
are expected to be sufficient.  In addition, working in 2D allows to neglect 
difficulties associated with the choice of eddy shape, which here is taken as circular.

The idea of mimicking a turbulent flow with a superposition of rigid vortices, whose dynamics
imitate as much as possible that of the real ones is not new \cite{synge43} (see \cite{abel00} 
for an application including the effect of Lagrangian sweep). In our case, 
thanks to the limited number of vortices per shell, this allows a drastic reduction in the 
number of degrees of freedom in the system. 
One more advantage is associated with the fact that a flow space structure is present and
incompressibility can be enforced. This prevents the difficulties present in more standard
Lagrangian stochastic models, which are associated with the need of a well mixed condition 
\cite{thomson87} imposed from the outside. The same problem occurs in the random walk 
model considered in \cite{kallio89,underwood93}, in which the vortices are modelled
as scattering centers for the particles.

The same approach based on the use of a superposition of vortices localized close to a
single particle can be utilized to analyze the separation of
particle pairs. The separation of trajectories in the two cases of a pair of fluid 
particles ($\tau_S=0$) and of a $\tau_S/\tau_L=0.26$ solid particle and a fluid one are
shown in Fig. 1 below.

\begin{figure}[hbpt]\centering
\centerline{
\psfig{figure=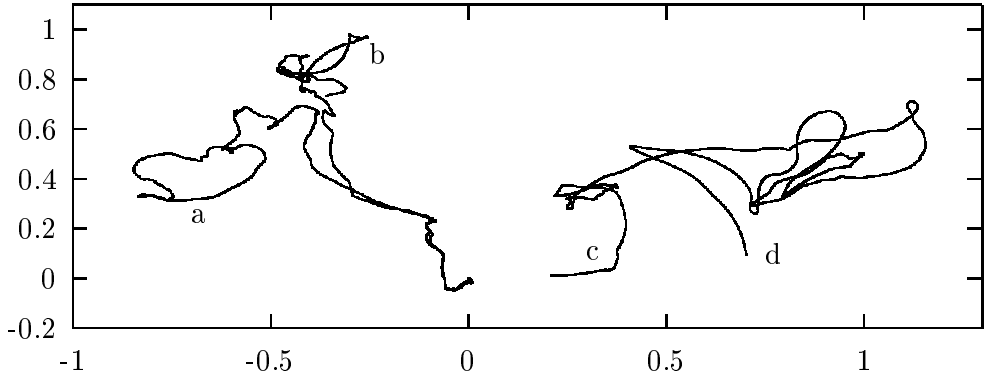,height=6.cm,angle=0.}
}
\caption{Two pairs of trajectories separating in a turbulent flow, in the absence of
external forces. 
Effective Reynolds number $L/\eta\simeq 10^5$ (17 shells). Trajectories $a$, $b$ and
$c$ refer to fluid particles; $d$ is the trajectory of a solid particle with $\tau_S/\tau_L=0.26$.
Notice the smoother path followed in $d$. Notice also the long transient in the first pair, 
where both particle are fluid and all the separation is due to fluid trajectory divergence.
}
\end{figure}
In the first case, an initial separation $0.25\eta$ was adopted; notice,
however, the long transient associated with the particles lying at ''viscous 
range'' separations.  The corresponding relative diffusion in velocity:
$\langle |\Delta u_i(t)|^2\rangle=\langle [u_i(t|\x_0,0)-u_i^\smalF(t|\x_0,0)]^2\rangle$ 
and in space
$\langle |\Delta x_i(t)|^2\rangle=\langle [x_i(t|\x_0,0)-x_i^\smalF(t|\x_0,0)]^2\rangle$ 
are shown in Figs. 2 and 3. 

\begin{figure}[hbpt]\centering
\centerline{
\psfig{figure=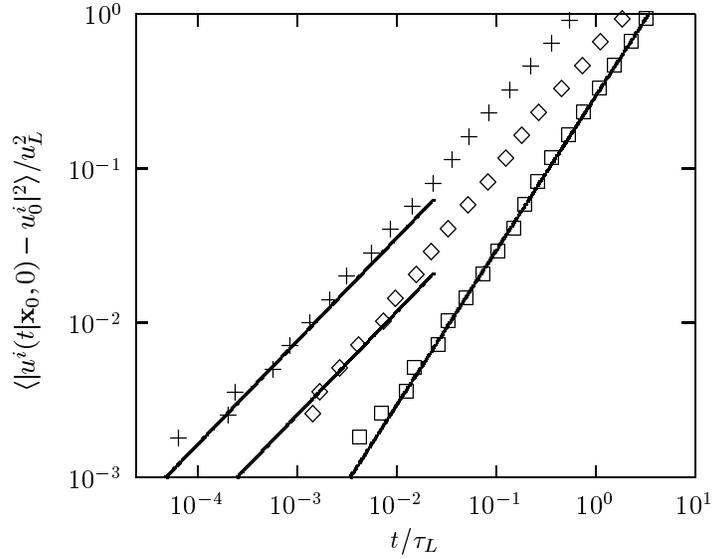,height=7.5cm,angle=0.}
}
\caption{Relative diffusion of velocity: $\Delta\u(t)=\u(t|\x_0,0)-\u^\smalF(t|\x_0,0)$;
effective Reynolds number $L/\eta\simeq 10^5$ (17 shells).
$\Box$: fluid particle diffusion ($\tau_S=\tau_G=0$), fitted with the linear law 
$\langle |\Delta u_i(t)|^2\rangle=0.29\frac{u_L^2t}{\tau_L}$. $\Diamond$: $\tau_S=0$, 
$\tau_G/\tau_L=0.01$;
$+$: $\tau_S/\tau_L=0.13$, $\tau_G=0$. The lower end of the data are fitted with $t^\frac{2}{3}$ 
slopes.
}
\end{figure}

\begin{figure}[hbpt]\centering
\centerline{
\psfig{figure=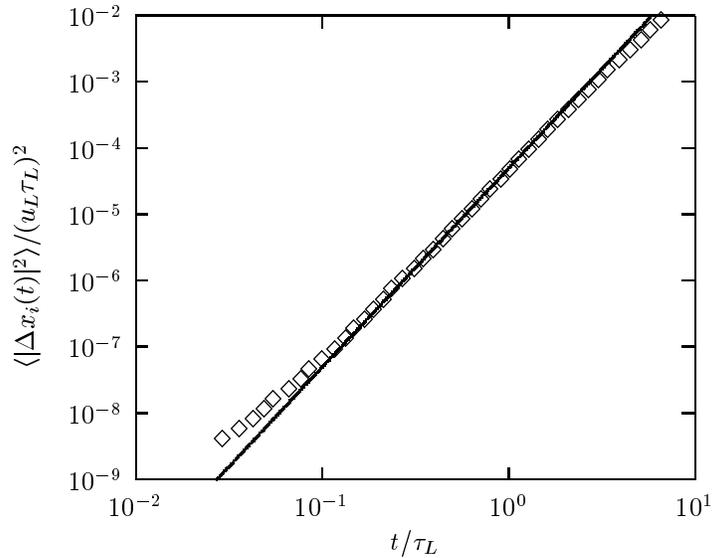,height=7.5cm,angle=0.}
}
\caption{Richardson diffusion of a pair of fluid particles and fit by 
$\langle |\Delta x_1(t)|^2\rangle=0.003\frac{u_L^2t^3}{\tau_L}$. 
Effective Reynolds number $L/\eta\simeq 10^5$ (17 shells).
}
\end{figure}
The parameter $\beta$ has been chosen to maximize the ratio $C_K^\frac{3}{2}/C_0$, and therefore 
the velocity variation due to trajectory separation. This choice of parameters resulted in:
$C_K^{-\frac{3}{2}}C_0\simeq 2.828$. In the artificial flow considered, the turbulent dissipation 
$\bar\epsilon$ is clearly undefined, and the constants $C_K$ and $C_0$ cannot be determined
individually. Assuming $C_K=2$, as in three dimensional turbulence, would give $C_0\simeq 8$, 
which is rather large, as are small the values of the constants $c$ and 
$\tilde c$ entering the expressions for velocity and coordinate relative diffusion 
$\langle |\Delta u_i(t)|^2\rangle=c\bar\epsilon t$ and 
$\langle |\Delta x_i(t)|^2\rangle=\tilde c\bar\epsilon t^3$. One finds:
$c/C_0\simeq 0.29$ and $\tilde c/C_0$, which would lead to: $c\simeq 2.3$ and 
$\tilde c\simeq 0.025$.  All of these constants are characterized by a strong scatter of 
experimental values; if the accepted value for $C_0$ is around 4 \cite{hanna81}, estimates for
$\tilde c$ went as low as $0.05$ \cite{fung92}, while recent measurements gave 
$\tilde c\simeq 0.25$ \cite{ott00}. The uncertainty in $c$ is magnified in $\tilde c$ due to 
the higher power of $t$, and a factor 10 in $\tilde c$ corresponds to a factor $\sim 2$ in 
$c$. Nonetheless, the small values of the ratios $c/C_0$ and $\tilde c/C_0$ 
(which, contrary to $c$ and $\tilde c$, are independent of $C_K$) indicate that, although the 
synthetic turbulence field satisfies 
the proper Lagrangian and Eulerian scaling, the ratio of relative diffusion to self-diffusion 
may be smaller than in real flows. We must not forget, however, that the relevance to three 
dimensional turbulence of parameters such as e.g. $C_K^\frac{3}{2}/C_0$, obtained from a two 
dimensional synthetic field, may be somewhat limited.

The analysis of trajectory separation, contrary to that of the self-diffusion 
$\langle |\u(t|x_0,0)-\u_0|^2\rangle$ is plagued  by several difficulties. The most important one
is the limitation in the sample size: small and large separations, in the relative diffusion
case, are calculated from the same sample of trajectory pairs, while, in the self-diffusion
case, the sample size is inversely proportional to the scale in exam. This limitation is 
particularly serious, since most of the information on the separation of solid particle
trajectories is conditioned to the initial velocity difference. The second difficulty is the 
strong effect of transients occurring when the particles are very close and stick together,
which is absent in the case of self-diffusion. In fact, as suggested in \cite{boffetta99}, 
to obtain the curves in Figs. 2 and 3, it was necessary to average over the logarithms of the 
doubling
times of $|\u(t|\x_0,0)-\u^\smalF(t|\x_0,0)|$ and $|\x(t|\x_0,0)-\x^\smalF(t|\x_0,0)|$, instead of
at fixed times, over $|\u(t|\x_0,0)-\u^\smalF(t|\x_0,0)|$ and $|\x(t|\x_0,0)-\x^\smalF(t|\x_0,0)|$.

The self-diffusion properties of particles with different values of the ratio $\tau_S/\tau_L$ are
shown in Figs. 4, and 5 for the two cases without and with external forces. 
\begin{figure}[hbpt]\centering
\centerline{
\psfig{figure=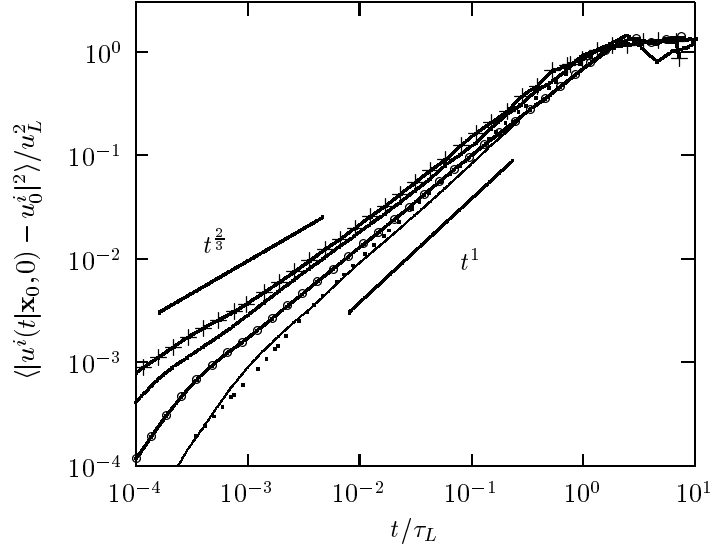,height=7.5cm,angle=0.}
}
\caption{Velocity self-diffusion in the inertia dominated case ($u_G=0$) 
for different values of $\tau_S/\tau_L$.
$'\,{\scriptstyle{-+-}}\,'$: 
$\tau_S/\tau_L=2.6$; heavy line: $\tau_S/\tau_L=0.26$; $'\,-\circ -\,'$: 
$\tau_S/\tau_L=0.026$; thin line: $\tau_S/\tau_L=0.0026$; dots: fluid particle. The Kolmogorov
time-scale is $\tau_\eta/\tau_L\simeq 3.8\times 10^{-4}$, corresponding to an effective Reynolds
number $Re\simeq 2\times 10^5$ (18 shells). In the various cases, we have, for $\sigma_{|\u-\v|}^2
\equiv\langle |\u-\v|^2\rangle-u_G^2$:
$\tau_S/\tau_L=0.0026$: $u_L^{-2}\sigma_{|\u-\v|}^2=0.0049$; 
$\tau_S/\tau_L=0.026$: $u_L^{-2}\sigma_{|\u-\v|}^2=0.047$; 
$\tau_S/\tau_L=0.26$: $u_L^{-2}\sigma_{|\u-\v|}^2=0.39$; 
$\tau_S/\tau_L=2.6$: $u_L^{-2}\sigma_{|\u-\v|}^2=0.9$. 
}
\end{figure}
\begin{figure}[hbpt]\centering
\centerline{
\psfig{figure=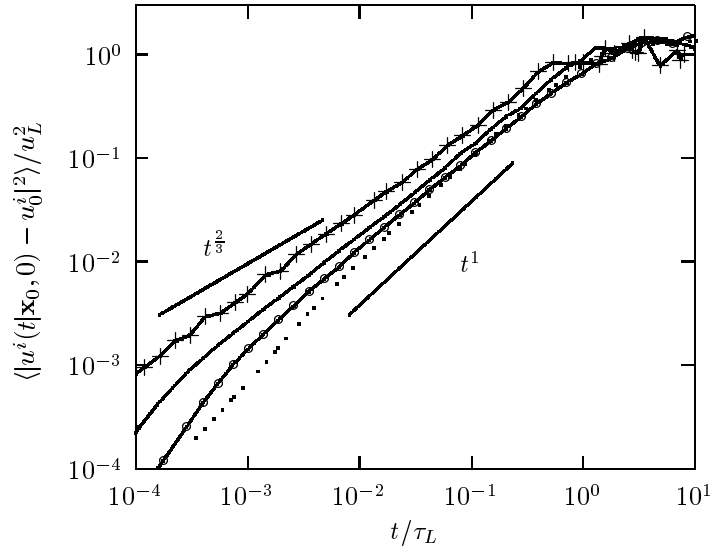,height=7.5cm,angle=0.}
}
\caption{Velocity self-diffusion in the external forcing dominated regime ($\tau_S=0$) for
different values of $u_G/u_L$; same flow parameters as in Fig. 4.
$'\,{\scriptstyle{-+-}}\,'$: 
$u_G/u_L=1.83$; heavy line: $u_G/u_L=0.09$; $'\,-\circ -\,'$: 
$u_G/u_L=0.01$; dots: unforced case. In the various cases we have:
$u_G/u_L=1.83$: $u_L^{-2}\sigma_{|\u-\v|}^2=1.7$;
$u_G/u_L=0.09$: $u_L^{-2}\sigma_{|\u-\v|}^2=0.11$;
$u_G/u_L=0.01$: $u_L^{-2}\sigma_{|\u-\v|}^2=0.015$;
}
\end{figure}

The time-scale of fluctuation for the difference $\u-\v$ appears to be
of the order of $\tau_G$ and $\tau_S$, respectively, in the forced and unforced case.
In the unforced case, the largest $\tau_S/\tau_L=2.6$ curve is basically indistinguishable 
from what would result from setting $\v=0$, i.e. $\tau_S=\infty$. (The short time behavior
of this curve, incidently, gives then also an idea of the applicability of the Taylor frozen 
field hypothesis in the absence of a mean flow). 
In the forced case, only the largest value of the drift $u_G$ leads to an appreciable
decrease in the correlation time for the fluid velocity sampled by the particle. In 
the other cases, in which $u_L>u_G$, the synthetic turbulence model is probably too crude
to give any definite answer.

The inertial range is compressed passing from space to time (compare with Fig. 11
in the Appendix) and this has the consequence that crossover behaviors 
dominate also at very high Reynolds number.  
In order to observe cleaner scaling behaviors in the velocity
self-diffusion, it would be necessary to condition the averages to the initial value 
of the difference $\u-\v$, as shown in Fig. 6 in the unforced case. 
\begin{figure}[hbpt]\centering
\centerline{
\psfig{figure=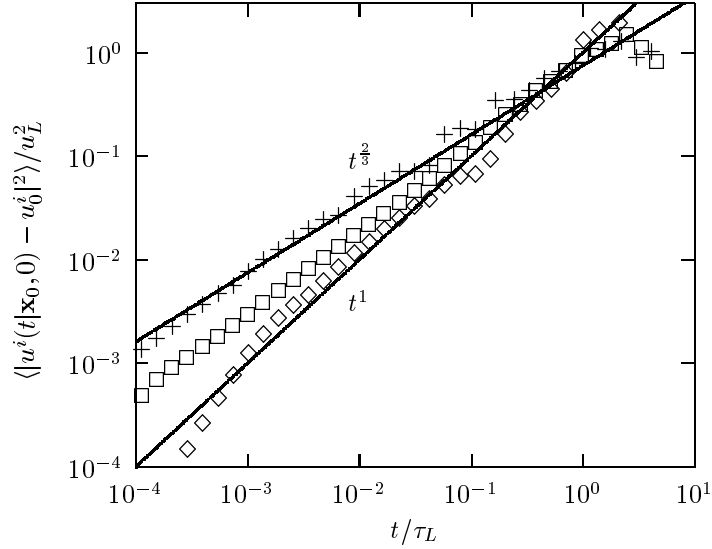,height=7.5cm,angle=0.}
}
\caption{Velocity self-diffusion conditioned to different values of the ratio 
$\langle|\u-\v|\rangle^{-1}|\u-\v|$, for fixed $\tau_S/\tau_L=0.26$, $u_G=0$. Same flow 
characteristics as in Fig. 4. 
$'\,{\scriptstyle{+}}\,'$: 
$2<\langle|\u-\v|\rangle^{-1}|\u-\v|$. 
$'\,\Box\,'$: 
$\langle|\u-\v|\rangle^{-1}|\u-\v|<0.12$.
$'\,\Diamond\,'$: unconditioned. Notice how ballistic behaviors are enhanced in 
the two conditioned cases.
}
\end{figure}

An example of cross-over behavior is the renormalization of 
the constant $C_0$ over the whole inertial
range for values of $\tau_S/\tau_L$ as small as $0.0026$. 

It is possible to have an idea of what is happening, at least in the unforced case, 
focusing on the separation of trajectories in a vortex of size $l>S$. The separation process 
is the result of the particle being accelerated by vortices of all the available scales
$l'$. From Eqn. (1), large vortices with $l'>l$ will produce
a constant velocity difference $u_{l'}\tau_S/\tau_{l'}$. The contribution to trajectory 
separation by a single vortex of size $l'$, will be, in the lifetime $\tau_l$:
$$
\Delta l(l')\sim\tau_lu_{l'}\tau_S/\tau_{l'}\sim l\tau_S/\tau_{l}(u_l/u_{l'}).
\eqno(16)
$$
The contribution from vortices at scale $l'\in[S,l]$, instead,
is similar to a random walk, since each vortex $l'$ produces an independent displacement of the
particle in the time $\tau_l$; the total displacement produced in this time will be therefore:
$$
\Delta l(l')\sim l'\tau_S/\tau_{l'}(\tau_l/\tau_{l'})^\frac{1}{2}\sim l\tau_S/\tau_l
\eqno(17)
$$ 
Smaller vortices, of size $l'<S$, produce altogether an $O(S)$ contribution to displacement.
Exploiting the fact that the vortices are distributed
uniformly in the logarithm of scale, and combining with Eqns. (16-17), the velocity shift
will be:
$$
\sum_{S<l'<l}\frac{\Delta l'u_{l'}}{l}\sim\sum_{S<l'<l}\sum_{S<l''<l'}
\frac{\Delta l'(l'')u_{l'}}{l}\sim[\log(l/S)]^2\frac{u_l\tau_S}{\tau_l}
\eqno(18)
$$
The effect on the time velocity structure function will be, from Eqn. (6):
$$
\langle |\u(t|\x_0,0)-\u_0|^2\rangle\sim C_0\bar\epsilon t(1+\frac{a\tau_S}{t}[\log(bt/\tau_S)]^2)
\eqno(19)
$$
with $a$ and $b$ constants. A fit of $\langle |\u(t|\x_0,0)-\u_0|^2\rangle$ using Eqn. (19) is shown
in Fig. 7. 
\begin{figure}[hbpt]\centering
\centerline{
\psfig{figure=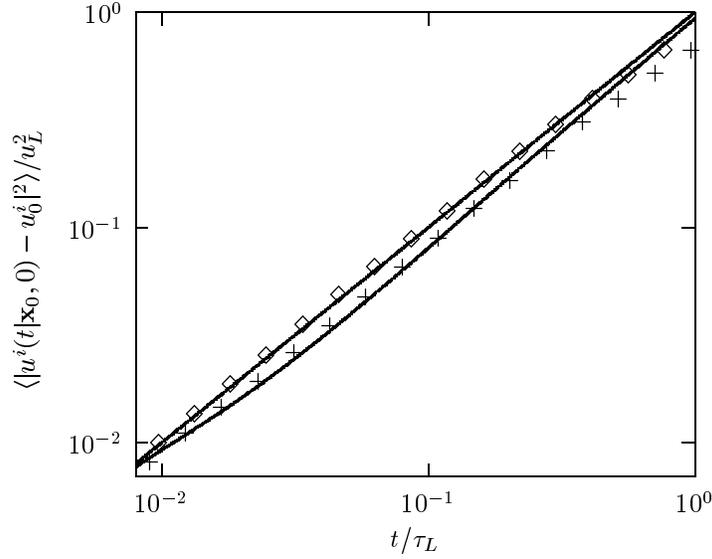,height=7.5cm,angle=0.}
}
\caption{Negative renormalization of $C_0$ for small values of $\tau_S/\tau_L$; $\Diamond$: 
fluid particle, fitted by Eqn. (14) with $C_0=8$ (for $C_K=2$). 
${\scriptstyle{+}}$: $\tau_S/\tau_L=0.0026$ $u_G=0$ fitted by Eqn. (11) with $C_0=8$, 
$a=-0.3$ and $b=1.4$.  Using still Eqn. (14), this would correspond to a renormalization 
$C_0\simeq 7$.
}
\end{figure}
It is to be noticed that the renormalization of $C_0$ is negative in all the inertia
dominated situations, with the exception of the $\tau_S\ge 1$ cases, in which the linear 
spectrum is essentially absent. A negative renormalization of $C_0$ corresponds to an 
inhibition of the time variation for $\u(t)$, which is akin to 
the increase of the correlation time for $\u(t)$, predicted
in \cite{pismen78,reeks77} in the case of a purely large scale flow.
It is not clear, however, how much these phenomena depend upon modelling assumptions.

The crucial aspect seems to be whether the time-statistics is given in an 
Eulerian frame or along trajectories. An Eulerian time-statistics would suggest
a faster fluid velocity variation in a Lagrangian frame \cite{kraichnan64}, which is 
the physical motivation of the results obtained in \cite{pismen78,reeks77}.
Following the same argument, a time-statistics given along trajectories would lead
to the fixed particles feeling the fastest velocity variation. 
Analytical calculations for random velocity fields with short correlation time and 
Lagrangian time statistics, however,
confirm the numerical result of slower velocity variations along particle trajectories
\cite{olla02}.


\vskip 10pt
\centerline{\bf IV. Application to Lagrangian transport modelling}
\vskip 5pt
The analysis carried on so far indicates that the strongest effect of particle inertia on 
velocity self-correlation is an enhancement of fluctuations at time-scales below
$\tau_S$ and $\tau_G$. Contributions at long time-scales, associated with renormalization of the
constant $C_0$, remain present even for very light particles, but the correction is small
compared to the uncertainties in the value of this constant, even in the case of fluid 
trajectories. Hence, for time scales above $\tau_S$ and $\tau_G$, it is acceptable to set
$\u\simeq\u^\smalF$ and to model the fluid velocity $\u$ sampled by the heavy particle, 
with the same Langevin equation used for $\u^\smalF$, i.e. the second of Eqn. (2). 
Being able to model the difference $\u-\u^\smalF$ becomes 
necessary, however, when $\tau_{S,G}\sim\tau_L$ and trajectory properties at time-scales
below $\tau_S$ and $\tau_G$ become important.

In a Markovian approach like that of \cite{desjonqueres88,berlemont90}, 
the equation for $\u$ would be obtained from the one for $\u^\smalF$, by adding, 
at discrete times $t_n=n\Delta$, $\Delta<\tau_{S,G}$, 
uncorrelated increments $\Delta\u$ extracted from a PDF (probability distribution function)
in the form $P(\Delta\u|\Delta\x,\u^\smalF)$, with $\Delta\x\sim (\v-\u)\Delta$ the trajectory
separation in the time $\Delta$. The shift $\Delta\u$ will have a mean part
$\Delta\bar\u(\u^\smalF)$, which is zero when $\u^\smalF=0$ and a
fluctuation 
$\Delta\tilde\u$: $\Delta\tilde u\sim u_{\Delta x}$ 
The cumulative effect of these increments would be felt also at time-scales above $\tau_S$ and 
$\tau_G$, in the form of additional noise and drift terms in the equation for $\u$. 
In particular, the amplitude of
the noise correction is estimated from $|\u-\v|\sim u_{S,G}$ and Eqns. (3,8,12):
$$
\Delta^{-1}(\Delta\tilde u)^2\sim\bar\epsilon (\tau_{S,G}/\Delta)^\frac{1}{3},
\eqno(20)
$$
which is an unphysical result: this amplitude remains finite also in the fluid particle limit 
$\tau_{S,G}/\tau_L\to 0$. The most the drift term can do to improve upon the situation, is to
make sure that $\langle u^2\rangle= u_L^2$ and that the only corrections be in the value of 
the constant $C_0$ and the Lagrangian correlation time $\tau_L$ seen by the solid particle. 
Neglecting the drift, which amounts to extracting the velocity increments from a PDF not 
conditioned to $\u^\smalF$, would lead to $\langle u^2\rangle >u_L^2$; this would be 
equivalent to say that solid particles privilege higher velocity regions of the flow, 
independently of the value of $\tau_{S,G}$. Going back to the discussion preceeding Eqn. (15),
neglecting the conditioning on $\u^\smalF$ would mean that any correlation among velocity
increments along fluid trajectories (longitudinal) and from trajectory separation (transverse),
is disregarded.  (It is to be mentioned that, contrary to 
the prediction of \cite{pismen78,reeks77}, the renormalization of the turbulent 
correlation time, necessary to the fluid velocity amplitude constant, leads
to a decrease rather than to an increment of $\tau_L$). Notice that approximating 
spatial correlations
with exponentials \cite{berlemont90,lu92} is not sufficient to cure this problem; the only
gain is eliminating the spurious noise dependence on the discretization $\Delta$.
In fact, if we neglected Kolmogorov scaling of space correlations and approximated
$\Delta\tilde u\sim u_A^2l/A$, we would still have $\Delta^{-1}(\Delta\tilde u)^2
\sim u_A^2/\tau_A\sim\bar\epsilon$, $A=S,G$, which remains finite when $\tau_{S,G}\to 0$ $[$compare 
with Eqn. (20)$]$.
In other words, anomalous diffusion is not an essential ingredient; memory
effects are.

Eliminating cumulative effects of the increments in $\u$ leads automatically to the
introduction of memory and to a non-Markovian model. A simple Gaussian model with these 
properties is obtained proceeding in the same philosophy of section III: the velocity
$\u$ is modelled by a superposition of solutions of stochastic linear equations, each associated 
with a spatial scale of the turbulent flow. For homogeneous isotropic turbulence (summation
over repeated vectorial indices is understood):
$$
\begin{cases}
u_i(t)=\sum_n\bar u_i^n(t);
\\
\frac{\d \bar u^n_i(t)}{\d t}-\bar\tau_n^{-1}A_i^n(\bar\u^n)=
\Big(\frac{c_n\bar\epsilon\tau_L}{\bar\tau_n}\Big)^\frac{1}{2}\xi_i(t)
\\
\langle\xi_i(t)\xi_j(0)\rangle=\delta_{ij}\bar\tau_n^{-1}\Xi(t/\bar\tau_n);
\qquad \int\d x\Xi(x)=1
\end{cases}
\eqno(21)
$$
where $A_i^n=-M_{ij}^n\bar u^n_j$, $\bar\tau_n$ and $M_{ij}^n$ are function of 
$\u-\v$,  
$c_n\simeq 2^{-\frac{n+2}{2}}C_0/2$, 
and for $|\u-\v|=0$: $\bar\tau_n=2^{-n}\bar\tau_0$. The noise $\xi_i(t)$ must have finite
correlation time in order to avoid that the dynamics of $\u$ at a given time-scale $\tau$ 
receive contribution 
by shells with $\bar\tau_n\gg\tau$ (this is implemented automatically in a numerical code, using
timesteps proportional to $\bar\tau_n$, in the different shells). With these prescriptions, for 
$|\u-\v|=0$, $\bar\u^n$ will fluctuate on time scale $\bar\tau_n$ with amplitude 
$\langle |\bar\u^n|^2\rangle=2^{-n}\langle |\bar\u^0|^2\rangle\propto\bar\tau_n$, and for 
$t\ll\tau_L$ Eqn. (15) will be satisfied.

The variable $\bar\u^n(t)$ in Eqn. (21) represents the contribution 
to the Lagrangian velocity component $\u(t)$ from vortices of size $l(n)$ such that 
$\tau_{l(n)}\sim\bar\tau_n$; hence 
$l(n)/L\sim 2^{-\frac{3n}{2}}$ and then also $\langle |\bar\u^n|^2\rangle\sim u_{l(n)}^2\sim 
u_L(l(n)/L)^\frac{2}{3}$. If $|\u-\v|\gg u_{l(n)}$, $\bar\u^n$ is the frozen velocity field of
the vortex of size $l(n)$. We have therefore:
$$
\frac{\bar\tau_n}{\bar\tau_n(0)}=
\begin{cases}
1 \quad &|\u-\v|\ll u_{l(n)}
\\
\gamma u_{l(n)}/|\u-\v|  &|\u-\v|\gg u_{l(n)}
\end{cases}
\eqno(22)
$$
where $\bar\tau_n(0)$ indicates the value of $\bar\tau_n$ for $|\u-\v|=0$ and $\gamma$ is 
an $O(1)$ constant. For 
$|\u-\v|\gg u_{l(n)}$, $\langle u_i^n(t)u_j^n(0)\rangle$ is just the space 
correlation function at separation $(\v-\u)t$ of the flow field produced by vortices at scale 
$l(n)$. Incompressibility fixes then its tensorial structure \cite{hinze}:
$\langle u_i^n(t)u_j^n(0)\rangle=f(t/\bar\tau_n)L_{ij}+g(t/\bar\tau_n)(\delta_{ij}-L_{ij})$ 
with $L_{ij}=\frac{(\u-\v)_i(\u-\v)_j}{|\u-\v|^2}$ and in three dimensions: 
$g(x)=f(x)+\frac{x}{2}f'(x)$.
Using Eqn. (13) also for $|\u-\v|\gg u_{l(n)}$ is equivalent to assuming exponential space 
correlations, and from $f(x)=\exp(-x)$ we find $g(x)= (1-\frac{x}{2})\exp(x)
\simeq\exp(-\frac{3}{2}x)$. In the opposite limit $|\u-\v|/u_{l(n)}\to 0$, isotropy imposes 
$\langle u_i^n(t)u_j^n(0)\rangle=f(t/\bar\tau_n)\delta_{ij}$. These conditions fix the form of 
the tensor $M^n_{ij}$ in the two opposite regimes:
$$
M^n_{ij}=
\begin{cases}
\delta_{ij}   & |\u-\v|\ll u_{l(n)}
\\
\frac{3}{2}\delta_{ij}-\frac{1}{2}L_{ij}
& |\u-\v|\gg u_{l(n)}
\end{cases}
\eqno(23)
$$
The model described by Eqns. (21-23) is reminiscent of a continuous time random walk 
\cite{montroll64}, in that the non-Markovian nature of the process originates from varying the
correlation times of the various contribution to $\u(t)$, rather than their amplitudes, which
remain invariant. 

The self-diffusion properties of the solution to Eqns. (21-23), limited to one dimension, are 
compared in Figs. 8 and 9, with the synthetic turbulence results of section III. The time $\tau_L$
and the noise amplitude $C_0\bar\epsilon$ in Eqn. (15) have been selected to fit the synthetic
turbulence result for $\tau_S=\tau_G=0$ and the expression
$\bar\tau_n/\bar\tau_n(0)=(1+0.3|\u-\v|/u_{l(n)})^{-1}$ has been used for the correlation times.
When $\bar\tau_n<\bar\tau_N(0)$, with $N$ the fastest shell, mode $n$ was disregarded and the 
fastest mode with correlation time above $2^{-N}\tau_L$ was renormalized to account for the 
energy of the lost shell.
\begin{figure}[hbpt]\centering
\centerline{
\psfig{figure=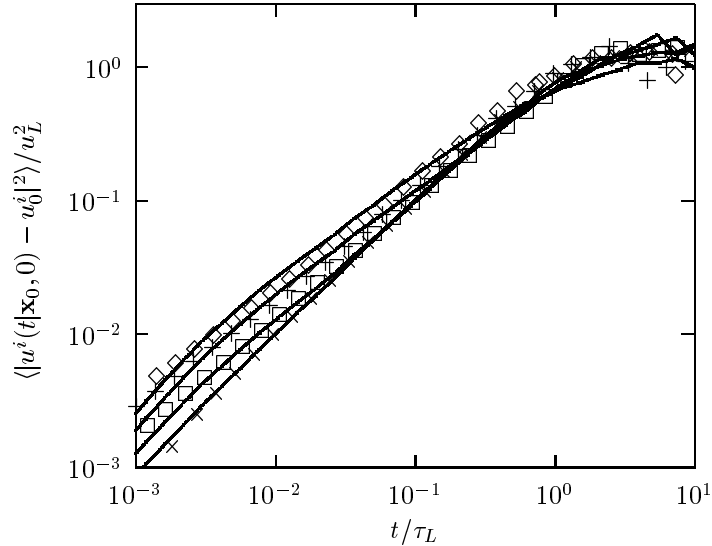,height=7.5cm,angle=0.}
}
\caption{Comparison of solution to 1-dimensional version of Eqns. (21-23) (continuous lines) and
synthetic turbulence results for velocity self-diffusion. $'\,\Diamond\,'$
$\tau_S/\tau_L=2.6$.
$'\,\scriptstyle{+}\,'$ $\tau_S/\tau_L=0.26$. $'\,\Box\,'$ $\tau_S/\tau_L=0.026$. $'\,\times\,'$
$\tau_S=0.$ Equation (21) has been solved with 13 modes and $\bar\tau_0(0)\simeq 0.35\tau_L$.
}
\end{figure}
\begin{figure}[hbpt]\centering
\centerline{
\psfig{figure=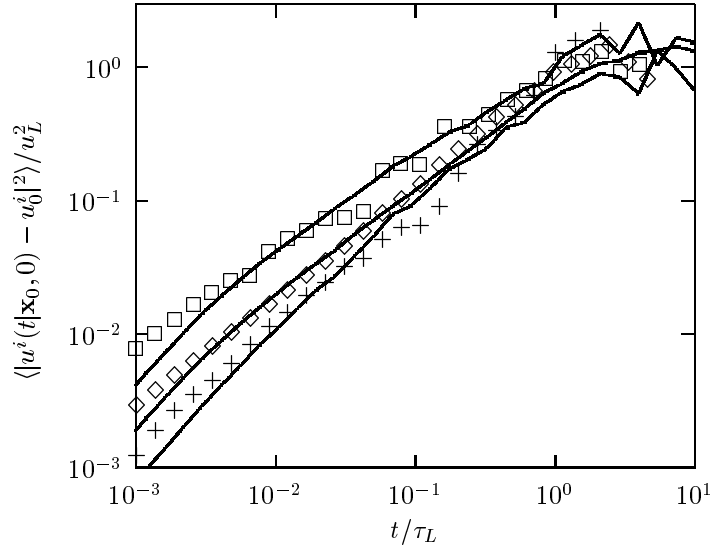,height=7.5cm,angle=0.}
}
\caption{Another comparison of solutions of Eqns. (21-23) (continuous line) and synthetic
turbulence results. Here, self-diffusion is conditioned to different values of the ratio
$\langle|\u-\v|\rangle^{-1}|\u-\v|$, for fixed $\tau_S/\tau_L=0.26$, $u_G=0$:
$'\,{\scriptstyle{+}}\,'$:
$2<\langle|\u-\v|\rangle^{-1}|\u-\v|$.
$'\,\Box\,'$:
$\langle|\u-\v|\rangle^{-1}|\u-\v|<0.12$.
$'\,\Diamond\,'$: unconditioned.
Equation (21) has been solved using
the same parameters of Fig. 8.
}
\end{figure}

The model just described allows to keep into account the Kolmogorov-like nature of
the small scale spatial structure of high Reynolds number turbulence. It is easy to see this.
From the discussion in
Section II, the typical particle-fluid relative velocity is $|\u-\v|\sim\max(u_S,u_G)$.
The fluid velocity increment along a solid particle trajectory $\langle |\u(t)-\u(0)|^2\rangle$
will be dominated by that shell for which $\bar\tau_n(u_{S,G}/u_{l(n)})=t$ and $t=\tau_{S,G}$ 
will give the cross-over from the situation in which the corresponding $\bar\tau_n$ obeys the 
first or the second of Eqn. (22).  
From Eqn. (22), and the scaling for the shell characteristic time and velocity:
$\bar\tau_n(0)=\bar\tau_02^{-n}$, $\langle |\bar\u^n|^2\rangle\sim u_L^2 2^{-n-1}$,
we thus obtain:
$$
\bar u_n/u_{S,G}\sim
\begin{cases}
(t/\tau_{S,G})^\frac{1}{2}
&\ t\gg\tau_{S,G}
\\
(t/\tau_{S,G})^\frac{1}{3}
&\ t\ll\tau_{S,G}
\\
\end{cases}
\eqno(24)
$$
At short times, the fluid velocity self-diffusion will have the $t^\frac{1}{3}$ scaling,
which comes from the spatial profile of the small vortices being seen as frozen by the crossing
solid particle. Conversely, at long times, we will find the usual $t^\frac{1}{2}$ normal scaling
of the Lagrangian time structure function $[$see Eqn. (4)$]$, 
which is obtained disregarding trajectory separation, and equating $t$ with the eddy turn-over 
time.
 
The problem of Markovian models \cite{desjonqueres88,berlemont90}, discussed at the beginning 
of this section, disappears in 
the present approach. Trajectory crossing effects, in the limit $\tau_{S,G}\to 0$,
are confined to shells with $n\to\infty$, and their contribution to $T_L$ disappears in 
the limit (as it should).
In this respect, this model is closer to the ones by \cite{sawford91,wang93}, in which 
the effect of particle inertia is accounted for by correlation time renormalization, rather
than modelling directly the change of fluid velocity from trajectory crossing. 
Of course, the intent of modelling the turbulent small scales, adopting either the Markovian 
approach of \cite{desjonqueres88,berlemont90}, or the non-Markovian approach used here and
in \cite{shao95,reynolds00}, was completely absent in \cite{sawford91,wang93}.

In basically all existing theories of heavy particle dispersion, the Kolmogorov structure
of small scale fluctuations is disregarded.  Typically, an
exponential \cite{berlemont90,lu92,wang93} or Gaussian \cite{pismen78,nir79} profile 
for the space correlations is adopted. In some approaches \cite{pismen78,nir79},
also the Lagrangian nature at short time separations, of the time correlations,
is neglected.

After first submission of the present work, this author became aware of the paper by Reynolds and
Cohen \cite{reynolds02}, in which a fractional Brownian motion algorithm was used to generate 
the anomalous scaling behaviors resulting from trajectory crossing. The advantage of the
present approach is the predictive ability for the scaling, whose physical mechanis is 
built in the algorithm by means of the shell superposition technique.
It is interesting to analyze the prediction of the present model on the test case considered
in \cite{reynolds02} based on the Snyder and Lumley experimental data \cite{snyder71}. 
Figure 10 illustrates a power law fit for the fluid velocity self diffusion along 
a solid particle trajectory simulated with Eqns. (21-23), with particle parameters corresponding
to the ''glass'', ''corn pollen'' and ''hollow glass'' particle case considered in \cite{snyder71}.
\begin{figure}[hbpt]\centering
\centerline{
\psfig{figure=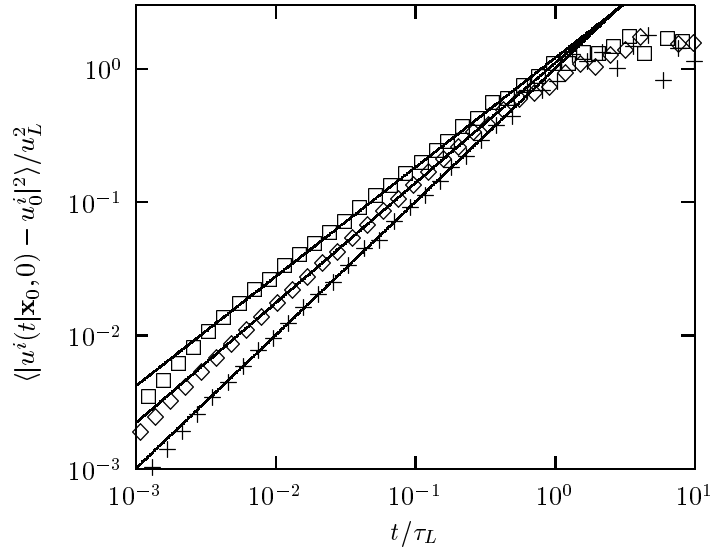,height=7.5cm,angle=0.}
}
\caption{Fluid velocity self-diffusion predicted by Eqns. (21-23) in three test cases;
$\Box$ (glass): $u_G/u_L\simeq 1.76$, $\tau_S/\tau_L\simeq 0.26$; $\Diamond$ (corn pollen)
$u_G/u_L\simeq 0.7$, $\tau_S/\tau_L\simeq 0.1$; ${\scriptstyle +}$ (hollow glass)
$u_G/u_L\simeq 0.052$, $\tau_S/\tau_L\simeq 0.01$. The continuous line illustrate the power
law fit by $t^{2H}$; in the three cases: $H\simeq 0.41$; $H=0.45$ and $H\simeq 0.5$.
}
\end{figure}
The same form for $\bar\tau_n$ used before: 
$\bar\tau_n/\bar\tau_n(0)=(1+0.3|\u-\v|/u_{l(n)})^{-1}$, leading to the results of Figs. 8 and 9,
continued to be used.
In the three cases the fit gives for the exponent $H$: $\langle |\u(t|\x_0,0)-\u_0|^2\rangle
\sim t^{2H}$, $H\simeq 0.41$,
$H\simeq 0.45$ and $H\simeq 0.5$. This agrees well with the results in \cite{reynolds02}
with the exception of the hollow glass case, in which a behavior closer to that of a fluid
parcel is predicted here.

A shortcome of the present model is clearly that the exact form of the shell time-scales
must be provided from the outside, since Eqn. (22) is only valid asymptotically. 
Among the other things,  knowing the exact form of $\bar\tau_n/\bar\tau_n(0)$ for 
$|\u-\v|/u_{l(n)}\lesssim 1$ would be necessary, at least for small $n$, to determine 
the fluid velocity correlation sampled by the solid particle.
This shortcome is common also to other models \cite{csanady63,wang93}, in which
the renormalization of the correlation time by inertia must be introduced as an external 
ingredient. It is to be stressed, however, that the inertia renormalization of the 
correlation time belongs, together with the determination of the constant $C_0$ \cite{sawford91a}
and of the relative size of Eulerian and Lagrangian correlation times \cite{hanna81},
to a class of problems which are havily influenced by the turbulent large scales. It is
thus to be expected that this parameter be dependent on the nature of the flow and it 
be necessary to provide it from the outside.

It is possible to extend the Lagrangian modelling approach illustrated in this paper to non-Gaussian
inhomogeneous turbulence situations, 
imposing, in the fluid particle limit $|\u-\v|=0$, individually on each shell, the well-mixedness 
condition \cite{thomson87}.
It must be stressed that the well mixedness condition can be imposed only in the fluid 
particle limit $\v=\u$. Apart of all the problems connected with ergodicity of solid
particle trajectories, the particle concentration profile in the inhomogeneous turbulence 
flow is unknown and, in fact, is typically one of the desired outputs of the model.

Basically, one has a condition on $\bar\epsilon(\x,t)$ from the integral time scale: 
$$
\bar\epsilon(\x,t)=\frac{u_L^2(\x,t)}{C_0\tau_L(\x,t)},
$$
plus a condition for each of the (tensor) moments which are experimentally available:
$$
\sum_n\langle [\bar\u^n]^p\rangle(\x,t)=\langle\u^p\rangle(\x,t)
$$
with Kolmogorov scaling giving, for $p=2$: $\langle \bar u_n^2\rangle(\x,t)=2^{-n-1}u_L^2(\x,t)$. 
The choice of the contribution from each shell to the non-Gaussian part of the statistics 
remains arbitrary, conditioned to statistical realizability of the moments for the $\bar u_n$.

A possibility could be to partition the contribution equally
among different shells: $\langle[\bar\u^n]^p\rangle_c(\x,t)=
2^{-\frac{p(n+1)}{2}}\langle[\u]^p\rangle_c(\x,t)$, with subscript $c$ staying for cumulant. 
A choice
which is more consistent with the standard practice in stochastic Lagrangian models, to 
let the non-Gaussianity be generated by a non-linearity in the drift term, considering it 
therefore to be a purely large scale feature of the flow, would be 
to concentrate all the non-Gaussianity production in the first shell:
$\langle|\bar\u^n|^p\rangle_c(\x,t)=\langle u^p\rangle_c(\x,t)\delta_{1,n}$. 

Once the PDF's $P_n(\bar\u^n,\x,t)$ for the various shells are obtained from the cumulants through
some reconstruction technique \cite{du94}, the Fokker-Planck equation for the second of Eqn. (21)
can be inverted to determine the drift terms $A^n_i$ from $\flux$ and $P_n$; considering for 
simplicity the case of stationary turbulence: 
$$
\partial_{\bar u_i^n}(A_i^nP_n)=-\bar\tau_n(0)\bar u_i^n\partial_iP_n
+\frac{c_n\flux\tau_L}{2}\partial_{\bar u^n_i}\partial_{\bar u^n_i}P_n
\eqno(25)
$$
As it is well known (see e.g. \cite{du94}), in inhomogeneous turbulence situations and for 
$\langle\bar\u^n\rangle=0$, Eqn. (25) leads to a constant contribution to the drift,
which counterbalances the drift $A^n_i$ from the gradient term $\bar u_i^n\partial_iP_n$ and
enforces zero mean for $\bar\u^n$. 

Notice that this balance disappears in the case of the solid particle flow, in which case the 
particle
velocity will be characterized by a non-zero turboforetic mean drift. The  multivariate
PDF $P(\{\bar\u^n\},\v,\x)$ corresponding to a uniform particle concentration
will obey the Fokker-Planck equation:
$$
v_i\partial_iP-\partial_{v_i}(\frac{v_i-u_i-u_{Gi}}{\tau_S}P)
+\sum_n\partial_{\bar u_i^n}(\bar\tau_n^{-1}A_i^nP)=
\sum_n\frac{c_n\flux\tau_L}{2}\partial_{\bar u^n_i}\partial_{\bar u^n_i}(P/\bar\tau_n)
\eqno(26)
$$
from which, for small $\tau_S/\tau_L$ (and then $u_S\ll\u_L$), the drift can easily be
calculated. Multiplying Eqn. (26) by $v_i$, integrating over $\v$ and $\{\bar\u^n\}$, 
and using $\langle v_iv_j\rangle=\langle u_iu_j\rangle+O(u_S^2)
\simeq\langle u_iu_j\rangle$, we obtain in fact:
$$
\langle v_i\rangle=-\tau_S\partial_j\langle u_iu_j\rangle
\eqno(27)
$$
Thanks to the well-mixedness condition, however, this turboforetic drift is free of 
spurious contributions from unproper Lagrangian modelling of the fluid limit case.

It must be said that, contrary to what happens (at least in part) in a model like \cite{thomson87},
the time structure of the individual trajectories is here physically relevant.  This because
of the coupling between $\u$ and $\v$ through Eqn. (1).
Trying to base an extension of the present model to non-homogeneous non-Gaussian turbulence,
on a statistical constraint like well-mixedness, may thus be of limited value.
It must also be said that the ergodic property is not satisfied in general by heavy 
particles. (In the case of fluid elements, the ergodic property is always assumed, implicitly, 
as a working hypothesis). It is not clear, therefore, whether it is appropriate to hypothesize 
invariance even only of the amplitude for $\u(t)$, when passing from $|\u-\v|=0$ to the heavy 
particle case $|\u-\v|>0$ \cite{olla02}. 
\vskip 10pt

\centerline{\bf V. Conclusions}
\vskip 5pt
The analysis carried out in  this paper provides a picture of the behavior of heavy particles 
in turbulent flows, and, in particular, a prediction for the scaling properties 
of the fluid velocity sampled by the particles, in function of the flow and particle 
properties. The predictions based on scaling analysis were confirmed by synthetic 
turbulence simulations and allowed the derivation of a Lagrangian transport model 
taking into account the non-Markovian nature of the fluid velocity along the heavy
particle trajectory. 

It is important to understand that this non-Markovian nature cannot be overlooked 
without getting into contradictions. For instance, it would be rather natural, in a model like 
\cite{desjonqueres88,berlemont90}, to hypothesize a normal diffusion behavior for $\u^\smalF$ at 
high Reynolds numbers, and to study pairs of particle-fluid trajectories of duration
$\Delta\le\tau_S$, in which the approximation $\u\simeq\u^\smalF$ is used.
But, as discussed at the beginning of Section IV, if one neglects correlations between the
velocity increments $\Delta\u=\u-\u^\smalF$ at the discrete times $t_n=n\Delta$, modifications 
in the time correlations sampled by the solid particle will appear up to time-scale $\tau_L$ and 
remain present in the fluid particle limit $\tau_{S,G}\to 0$, which is unphysical. Even worse, 
if the PDF $P(\Delta\u|\u^\smalF)$, from which $\Delta\u$ is extracted, is not chosen
carefully enough, and the condition on $\u^\smalF$ is disregarded by choosing say
$P(\Delta\u)=P(\u^\smalE(\x+\Delta\x,t)-\u^\smalE(\x,t))$, one will find also $\langle u_i^2\rangle
\ne u_L^2$.

The scaling analysis suggests a picture of transients between 
different spectral slopes at different time-scales, in which the controlling parameters
are $\tau_S$ and $u_S$ ($\tau_G$ and $u_G$ in the case of strong external forces), which are the
time and velocity scales for $\u-\v$. Both synthetic turbulence simulations
and the runs from the Lagrangian model, however, indicated that cross-over behaviors rather than
pure scaling dominate even for very high Reynolds numbers.

In the last section, some ideas for inclusion of non-Markovian effects in a Lagrangian
stochastic model for solid particle transport have been illustrated.
In this way, the Kolmogorov nature of the turbulent
small scales and the Lagrangian character of time correlations is fully taken into
account. 
The suggested approach is based on the use of a superposition of Markovian 
processes, which will be slower than a simpler Markovian model by a factor of the order
of the number of terms in the superposition.  (In our case, with 12 shells, the 
ratio was approximately 8). 
A similar approach has been recently used to model turbulence 
intermittency \cite{biferale97}. 

An alternative approach to generate the sub-diffusive
behavior for the velocity, at time-scales below $\tau_{S,G}$,
could have been to resort to a continuous
time random walk \cite{montroll64}, or, equivalently, to L\`evy walks \cite{schlesinger87}.
This would have been lighter from the computational point of view, but the control on the
form of the PDF for the velocity and the velocity difference would have been much more difficult.
The control on the shape of the correlations for long times, would have been difficult as well.
Notice that, in the present approach, recirculation phenomena and correlations modelled on
Frenkiel functions \cite{hinze,desjonqueres88} could easily have been taken into account
in Eqn. (21) by the use of complex relaxation times. 

Compared to the recent model by Reynolds and Cohen 
\cite{reynolds02}, the present approach provides in self-consistent way the scaling
exponent for the fluid velocity self-diffusion. This both in the case of dominant inertia
and dominant external forces and with no special cost in terms of simplicity and of the
possibility of imposing a well mixedness condition in inhomogeneous non-Gaussian turbulence
situations.

To obtain a Lagrangian model of practical use, a precise form for the dependence of the shell 
correlation time $\bar\tau_n$ on $\u-\v$ should be provided. The two-dimensional synthetic 
turbulence simulations, which have been carried out, can at most confirm scaling behaviors
and perhaps give the sign of the inertia produced correction to the correlation time $[$see
Eqn. (19)$]$. The form of the correlation times $\bar\tau_n$ could probably be obtained only by 
DNS of the full Navier-Stokes equation in the proper tree-dimensional flow conditions.

However, although other such attempts to take into account the Kolmogorov structure of the 
turbulent small scales, were already carried on \cite{shao95,reynolds00}, it is still not clear 
how relevant these aspects are to the derivation of a dispersion model of practical use. 
Probably, the main interest remains conceptual.
Considering the reduced information on the turbulent flows available in practical situations,
and the fact that dispersion is governed by the local large scales anyway, a minimal 
implementation of the model could disregard the small scales altogether and it could be 
sufficient to keep only one 
shell in Eqn. (21). In such a hyper-simplified version, the present model would become 
equivalent to \cite{wang93}, with a generalized form of the Lagrangian correlation profile
for $\tau_S=0$ to allow enforcement of the well mixedness condition.

\vskip 5pt
\noindent{\bf Aknowledgements:} I wish to thank Paolo Paradisi for interesting discussion and
for his contribution to the initial part of this research, and Francesco Tampieri for insightful
comments. 

\vskip 10pt
\centerline{\bf A. Appendix}
\vskip 5pt
In this appendix, a closer look is given to the synthetic turbulence model described in section
III. In this approach, the effect of a homogeneous turbulent flow is modelled by a velocity field 
localized around the particles that one is tracking. This limits the evaluation of spatial
objects like e.g. correlation functions, to the case in which the correlation points coincide
with the particles. This approach, has been used to analyze the process of relative diffusion
(see Figs. 2 and 3), but, as already discussed, is plagued by difficulties. 
The space structure functions can be obtained in a simpler way from the assumption that the
local field around the particle is representative of the whole flow. 
The Eulerian velocity field is in the form:
$$
\u^\smalE(\x,t)=\sum_i\bar\u_i(\x,t)
\eqno({\rm A}1)
$$
where $i$ labels the vortex. The contribution from vortex $i$ is 
$$
\bar\u_i(\x,t)=
\alpha r_i^\frac{1}{3}W_i{\bf R}(\frac{\x-\bar\x_i(t)}{r_i})T(\frac{t-\bar t_i}{\bar\tau_{r_i}})
\eqno({\rm A}2)
$$
where $W_i$ are uncorrelated zero mean and unitary variance random variables, $\alpha$ is a 
constant and $\bar\x_i(t)$ and $\bar t_i$ indicate respectively the vortex center position 
at time $t$ and
the time of ''birth'' of vortex $i$. We have considered vortices with compact support in space and 
time; precisely:
$$
\begin{cases}
{\bf R}(\x)=\e_3\times\x H(1-x)
\\
T(t)=H(t(1-t))
\end{cases}
\eqno({\rm A}3)
$$
where $H(x)=x\theta(x)$ with $\theta(x)$ the Heaviside step function. The lifetime of the vortex 
has been taken as a deterministic function of the radius: $\bar\tau_r=\beta r^\frac{2}{3}$.

Since the $W_i$ are uncorrelated, the
two-point correlation reduces to a sum of contributions from the individual vortices:
$$
\langle\u(\x,t)\u(0,0)\rangle=\alpha^2\int\d\log r\int\d\bar t\int\d^2\bar x N(r)r^\frac{2}{3}
T(\frac{t-\bar t}{\bar\tau_r})T(\frac{-\bar t}{\bar\tau_r}){\bf R}(\frac{\x-\bar\x}{r})
{\bf R}(-\frac{\x}{r})
\eqno({\rm A}4)
$$
If any point feels on the average the effect of a fixed number of vortices per shell, the vortex 
density $N(r)$ can be written in the form 
$$
N(r)=\bar N(\pi r^2\bar\tau_r)^{-1}
\eqno({\rm A}5)
$$
with $\bar N$ a constant, which is exactly $1/\log 2$ in our case of one vortex per octave
contributing to velocity at a given time and position.

Substituting Eqns. (A3) and (A5) into
Eqn. (A5), one can find the expression for the longitudinal strucure function
$S_2(l)=\langle (u_1(l\e_1,0)-u_1(0,0))^2\rangle$:
$$
S_2(l)=\frac{\alpha^2}{15\pi\ln 2}\int\frac{\d r}{r^\frac{1}{3}}\int\d^2\bar x
R_1(\bar\x)[R_1(\bar\x+(l/r)\e_1)-R_1(\bar\x)] 
\eqno({\rm A}6)
$$
Integrating numerically, one finds: $S_2(l)\simeq 0.0039\alpha^2 l^\frac{2}{3}$, i.e. in terms
of Kolmogorov constant and mean dissipation:
$$
\alpha\simeq 16C_K^\frac{1}{2}\bar\epsilon^\frac{1}{3}
\eqno({\rm A}7)
$$ 
In Fig. 11 below is illustrated the structure function for an 18-shell synthetic
tubulence field.
\begin{figure}[hbpt]\centering
\centerline{
\psfig{figure=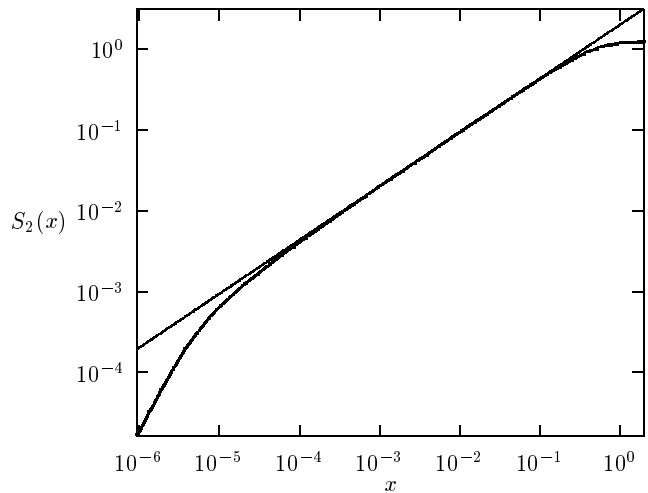,height=7.5cm,angle=0.}
}
\caption{Space structure function from an 18-shell synthetic turbulence field with $\alpha=22.627$,
$L=1$ and fit by a Kolmogorov spectrum $S_2(x)=2x^\frac{2}{3}$.
}
\end{figure}
All simulations have been carried out with $\beta\simeq 8.3/\alpha$, corresponding to an eddy life
of the same order of the eddy turn-over time. (Large and small values of $\beta$ 
correspond respectively to long and short vortex lives). 

Once we know the coefficients for 
$x^{-\frac{2}{3}}S_2(x)$ and $t^{-1}\langle|u_1(t)-u_1(0)|^2\rangle$, from Eqns. (10) and (A7), we
can eliminate the dissipation $\bar\epsilon$ to find the ratio $C_0^{-\frac{3}{2}}C_k$. We obtain
from the simulations $C_K^{-\frac{3}{2}}C_0\simeq 2.828$. If we take for the Kolmogorov constant
$C_K\simeq 2$, we will have $C_0\simeq 8$, which is somewhat large. The explanation of this fact may
lie in the way a particle deviates from the flow line of a given vortex due to the effect of 
vortices at other scales in the same region. In such a model, this effect is magnified by the 
rigid structure of the
vortices, and a particle will move out of the one in which it lies at a given time, faster than 
it would do if the vortex were able to be deformed by the flow. More important, however, is
probably the fact the this flow is two-dimensional and it would be interesting therefore to
compare it with a closer physical situation like two-dimensional turbulence in the inverse cascade 
range.

Following the same approach, it is possible to calculate $u_L$ and, from Eqns. (3-4) also
$L$ and $\tau_L$. The main parameters of the synthetic turbulent field are listed in Table I
below.
\vskip 20pt
\centerline{
\begin{tabular}{|p{2cm}|p{2.5cm}|p{2.5cm}|}
\hline
$u_L\simeq 0.61$ & $T_L\simeq 0.07$ & $C_0^{-1}\tilde c \simeq 0.003$ \\
$L\simeq 0.17$ & $C_K^{-\frac{2}{3}}C_0\simeq 2.83$ & $\alpha=22.627$ \\
$\tau_L \simeq 0.2$ & $ C_0^{-1}c\simeq 0.29$ & $\beta=0.367$ \\
\hline
\end{tabular}
}
\vskip 10pt
\centerline{Table I}

The last issue is how the vortices were generated in the synthetic turbulence field. The total 
number of vortices in shell $n$ was taken equal to the integer part of a variable ${\cal N}$, 
obeying an equation in the form:
$$
\dot{\cal N}_n=-\tau_n^{-1}(|\u-\v|){\rm sign}(N_n-1)
\eqno({\rm A}8)
$$
where $N_n$ is the number of vortices having the particle in their support at the given time
and $\tau_n(|\u-\v|)$ is the average time a particle travelling across the fluid with relative
velocity $\v-\u$ feels the effect of a vortex in shell $n$:
$$
\tau_n(|\u-\v|)\sim 2^{-\frac{2n}{3}}F(2^{-\frac{1}{3}}|\u-\v|)
\eqno({\rm A}9)
$$
with $F(x)\propto x^{-1}$ for $x$ large.
In this way, the number of vortices could be adjusted, in a time of the order of the vortex life,
in such a way that the particle felt the effect of only one vortex per shell. 

If the number of vortices alive at a given time was less than ${\cal N}$, a new one was put in 
the shell in an appropriate domain around the particle and with an age chosen at random between
zero and the vortex lifetime. The domain was an annular region comprised between $r$ and $\gamma r$
with $\gamma>1$, in such a way that the velocity at the particle position did not suffer from 
discontinuities. In all simulations, we set $\gamma=2.3$; with this choice, it was enough to
keep ${\cal N}\le 4$ in all simulations, with the exception of those with $\tau_S/\tau_L=2.6$
which required ${\cal N}\le 5$. As in other shell models, the main limitation was not the
number of degrees of freedom, but the strongly different time scales involved when the number
of shells becomes large.

Apart of the potential increase in the number of eddies required in the neighborhood of
a particle, and of the conceptual problems associated with the eddy geometry, extension of
the present technique to three dimensions is clearly straightforward.


\end{document}